# Ultra-short-pulse high-average-power Megahertz-repetition-rate coherent extreme-ultraviolet light source

R. Klas[1,2,*], A. Kirsche[1,2], M. Gebhardt[1,2], J. Buldt[1], H. Stark[1], S. Hädrich[3], J. Rothhardt[1,2,4], J. Limpert[1,2,4]

[1]*Institute of Applied Physics, Abbe Center of Photonics, Friedrich-Schiller-University Jena, Albert-Einstein-Str. 15, 07745 Jena, Germany*

[2]*Helmholtz-Institute Jena, Fröbelstieg 3, 07743 Jena, Germany*

[3]*Active Fiber Systems GmbH, Ernst-Ruska-Ring 17, 07745 Jena, Germany*

[4]*Fraunhofer Institute for Applied Optics and Precision Engineering, Albert-Einstein-Str. 7, 07745 Jena, Germany*

*Corresponding author: robert.klas@uni-jena.de

**High harmonic generation (HHG) enables coherent extreme-ultraviolet (XUV) radiation with ultra-short pulse duration in a table-top setup. This has already enabled a plethora of applications. Nearly all of these applications would benefit from a high photon flux to increase the signal-to-noise ratio and decrease measurement times. In addition, shortest pulses are desired to investigate fastest dynamics in fields as diverse as physics, biology, chemistry and material sciences. In this work, the up-to-date most powerful table-top XUV source with 12.9 mW in a single harmonic line at 26.5 eV is demonstrated via HHG of a frequency-doubled and post-compressed fibre laser. At the same time sub-6 fs XUV pulse duration allows accessing ultrafast dynamics with an order of magnitude higher photon flux than previously demonstrated. This concept will greatly advance and facilitate applications of XUV radiation in science and technology and enable photon-hungry ultrafast studies.**

## Introduction

Since the first experimental demonstration of high harmonic generation in the late 1980s[1,2], strong efforts have been made to enhance the average power of laser-like sources in the XUV[3], enabling applications on the atomic length-(nanometer)[4] and time-scale (femtosecond to attosecond)[5,6]. In the early stages up to the year 2010 (Fig. 1), Ti:Sa based amplifiers have been proven as an effective driver for HHG[3,7,8], since they provide ultra-short pulse durations (~25 fs) at 800 nm wavelength and high pulse energies (several millijoule). However, their limited average power of a few tens of watts in best case[9], limited the XUV flux of such systems to sub-100 μW per harmonic line (Fig. 1). An increase in XUV average power would help for example to mitigate space charge effects in photoelectron emission spectroscopy[10] (at high repetition rates), as well as to shorten acquisition times and, hence, enhance the signal-to-noise ratio in (time-resolved) coincidence measurements[11], XUV-absorption spectroscopy[12], XUV-ionization spectroscopy[13], coherent diffractive imaging of ultrafast magnetization dynamics[14], fluorescence spectroscopy[15] and XUV-pump XUV-probe experiments[16,17], among others. Furthermore, shortest pulses are desired to investigate fastest dynamics in atoms[6,18], molecules[19], ions[16], solids[12] and compound materials[20].

Inherently, an increase in XUV average power can be achieved with high average power driving lasers. In recent years, enhancement cavities[21,22] as well as Yb-based fibre lasers[23], which are capable of much higher average powers in the kilowatt regime (enabling megahertz repetition rates[24]), have shown a first increase in XUV average power demonstrating >100 μW in a single harmonic line[22,23], with typical conversion efficiencies $< 10^{-5}$. Further increase of the HHG efficiency and, hence, XUV average powers in the milliwatt-regime can be achieved by using a combination of an optimized enhancement cavity together with a high average power driving laser[25], or by using short wavelength drivers[26,27]. The latter ones made use of the efficiency scaling of the single atom response with $\lambda^6$, where $\lambda$ represents the driving wavelength[28], increasing the HHG efficiency by more than one order of magnitude.

Shorter driving pulses can enhance the efficiency and the cutoff energy of the XUV comb even more and naturally generate shorter pulses in the XUV. This can be understood since higher intensities $I$ can be applied for phase-matched HHG[3]. Furthermore, the single atom dipole amplitude $A_q$ scales with $I^{4.5}$ [29]. Since the macroscopic yield scales with $A_q^2$ [7], the HHG efficiency scales with $I^9$, resulting in higher efficiencies for shorter pulses and a ~3 times shorter XUV pulse duration[30].

However, two-photon absorption in high-reflective and anti-reflective coatings as well as in transmission optics, and the consequential heating of the materials, makes it very challenging to generate high average power as well as ultra-short pulse duration driving lasers with <550 nm wavelength. Therefore, until now, short wavelength driven HHG with milliwatt XUV average power was done using >85 fs pulses and average powers of a few Watt[25,26].

In this letter, HHG driven by the unique combination of high average power (51 W), short driving wavelength (515 nm) and ultra-short pulse duration (18.6 fs) is demonstrated. This ultrashort visible pulses enable the generation of sub-6 fs XUV pulses with a conversion efficiency of $2.5 \cdot 10^{-4}$, and, consequently the high average power of the driving laser boosts the photon flux to a record of 12.9 mW in a single harmonic line at 26.5 eV - surpassing previously demonstrated sources by an order of magnitude (Fig. 1).

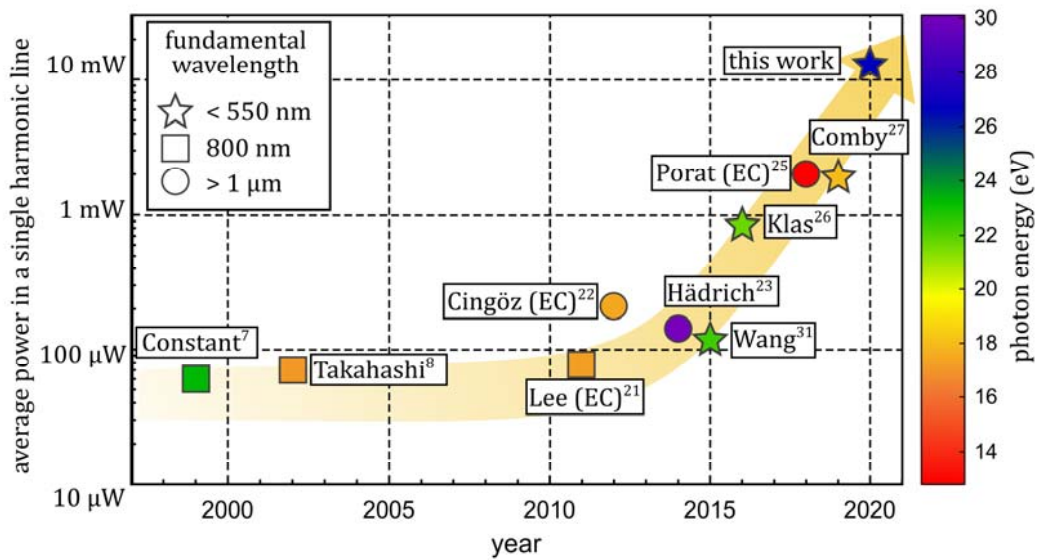

**Fig. 1 State of the art high harmonic sources.** Assorted sources with, at the time, record high average powers above 10 μW in a single harmonic line for photon energies ranging from 12 eV to 30 eV over a time span from 1999 until to date[7,8,21–23,25–27,31]. The photon energy is indicated by the marker colour, while the driving wavelength is indicated by the marker shape. Squares are for Ti:Sa based systems with a wavelength of 800 nm, circles are for Yb-based systems at >1 μm wavelength and stars are for cascaded schemes for high harmonic generation with the second or third harmonic of the prior mentioned architectures. (EC) marks sources based on an enhancement cavity, while all other sources are in a single pass geometry.

## Results

**Pulse compression at high average power and 515 nm wavelength**

For the experiment performed herein, a 515 nm laser at a repetition rate of 1 MHz, 89 W of average power and 200 fs is compressed, using a nonlinear hollow core fibre compressor. Careful selection of UV-grade fused silica glasses and coating materials for transmission and reflection optics, that have a low two-photon absorption and a large OH-content of up to 1000 ppm, make it possible to handle such high average powers at 515 nm. Simulations to optimize the output peak power while maintaining a low ionization level, revealed an optimal fibre diameter of 150 μm at a fixed length of 1 m. Spectral broadening, shown in Fig. 2 a), is achieved by filling such a capillary with 0.8 bar krypton. The transmission of 59 % is the same for an evacuated as well as a gas-filled capillary, showing that no significant ionization is present. Temporal compression of the spectrally broadened pulses is done using a custom designed chirped mirror compressor, which shows virtually no heating of the chirped mirrors due to their high reflectivity, resulting in a transmission of 96 %. The overall dispersion is - 2000 fs² (20 bounces on -100 fs²/bounce mirrors), which is also compensating the glass dispersion of the used lenses and windows of 1200 fs². The broadened spectrum is used to calculate the corresponding deconvolution factor, uncovering a compressed pulse duration of 18.6 fs (Fourier limit 16.5 fs) at a record average power of 51 W and a pulse energy of 51 µJ, resulting in a peak power of >2.5 GW. Furthermore, a spatial characterization of this compressed beam shows a nearly diffraction limited beam quality with an M² value of 1.25 x 1.26. This unique combination of short wavelength, ultrashort pulse duration, high average power and very good beam quality represents a novel class of driving laser for high harmonic generation.

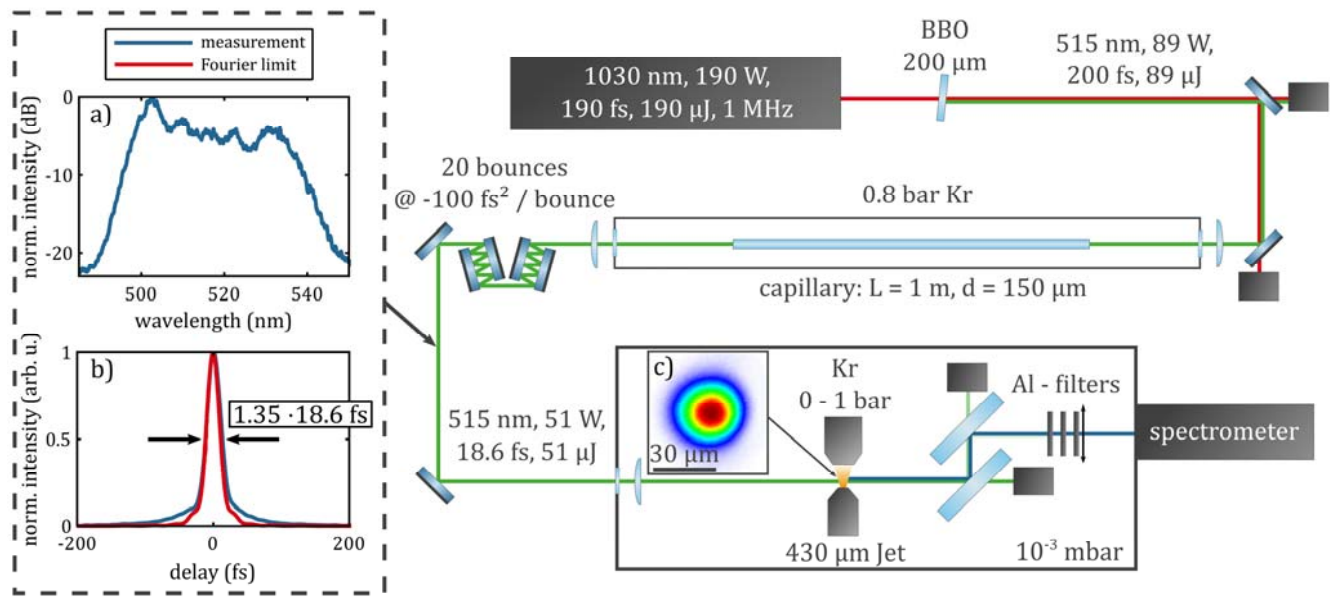

**Fig. 2 Setup for efficient high average power HHG.** The Yb-fibre laser is frequency doubled in a BBO crystal. Afterwards, the 515 nm light is coupled into a gas-filled capillary for spectral broadening and is subsequently compressed using a chirped mirror compressor. The spectral and temporal pulse characteristic at 51 W of average power is shown in the insets **a)** and **b)**. HHG is achieved by focussing the short visible pulses to a diameter of 33 µm (1/e² intensity,**c)**) inot a krypton gas jet. Separation of the driving laser and the XUV light, as well as additional attenuation of the XUV radiation (in order not to saturate the detector) is done using two glass plates in Brewster's angle and additional aluminium filters. At the end, the XUV beam is characterized spectrally and spatially using a flat field spectrometer equipped with an XUV CCD camera.

**High harmonic generation results**

HHG is achieved by focussing these pulses to a diameter of 33 µm (1/e² intensity, Fig. 2c)) into a krypton gas jet. The generated harmonics and the remaining visible light pass through two glass plates in Brewster's angle and various 1 µm thick aluminium filters, not only to separate the higher average power driving laser from the generated extreme ultraviolet (XUV) light, but also to attenuate the XUV light for further analysis. Subsequently, a flat field spectrometer equipped with a CCD camera is used to analyse the XUV beam spectrally and spatially. Optimization of the XUV flux is done by iteratively optimizing the nozzle position on a 3D-translation stage and an iris in front of the vacuum chamber. The optimal phase-matching pressure is found via variation of the applied pressure, revealing an optimum at 0.4 bar. The result of a corresponding simulation with a one dimensional model support these findings, assuming a distance between the laser beam and the nozzle of 144 µm. This results in a particle density in the interaction region of $3.8 \cdot 10^{18}$ cm⁻³ and an absorption length of 130 µm at 26.5 eV. Consequently, the medium length defined by the nozzle diameter (430 µm), which is much shorter than the Rayleigh length (1.3 mm), allows for absorption-limited HHG [7].

Due to the ultra-short pulse duration at 515 nm, the HHG efficiency into a single harmonic line at 26.5 eV is $2.5 \cdot 10^{-4}$, which is among the highest reported so far[8,27]. The short pulse duration is important for this high efficiency since higher intensities can be applied for phase-matched HHG, compared to previous experiments using 85 fs driving laser pulses[26], the shorter driving laser pulses show a threefold increase in efficiency as well as a >5 eV higher cut-off energy. Furthermore, the high average power (51 W) of the driving laser allows 23.1 mW of average power in the range from 20 eV to 35 eV (Fig. 3 a)), with a record high average power of 12.9 mW in the strongest harmonic line at 26.5 eV. In addition, the increased cutoff allows for >1 mW of average power above 30 eV. Note, that for a high flux delivery, a gracing incidence plate[32] or annular beam driven HHG[33] in combination with a 100 nm aluminium filter could be used as a separator, resulting in an attenuation of the driving laser to <1 μW with >5 mW usable XUV average power on target. The Fourier-limited pulse duration of a single harmonic line corresponds to 3.4 fs pulse duration in the XUV (Fig. 3 b)), which is in good agreement with the duration of the simulated phase-matching window. Focussing these pulses to a spot size of 1 μm, would result in an intensity $> 10^{14}$ W/cm$^2$. Thus, even nonlinear XUV-techniques are in reach at a MHz repetition rate. A long-term stability measurement of the photon flux of the strongest harmonic line at 26.5 eV shows an rms-deviation of 2.6 % over a time period of 30 min (Fig. 3 c)). Furthermore, a spatial lineout integrated over the whole spectrum together with a Gaussian fit shows a Gaussian like beam profile of the XUV beam with a divergence of 2 mrad (Fig. 3 d)), showing the great potential of this source for further experiments.

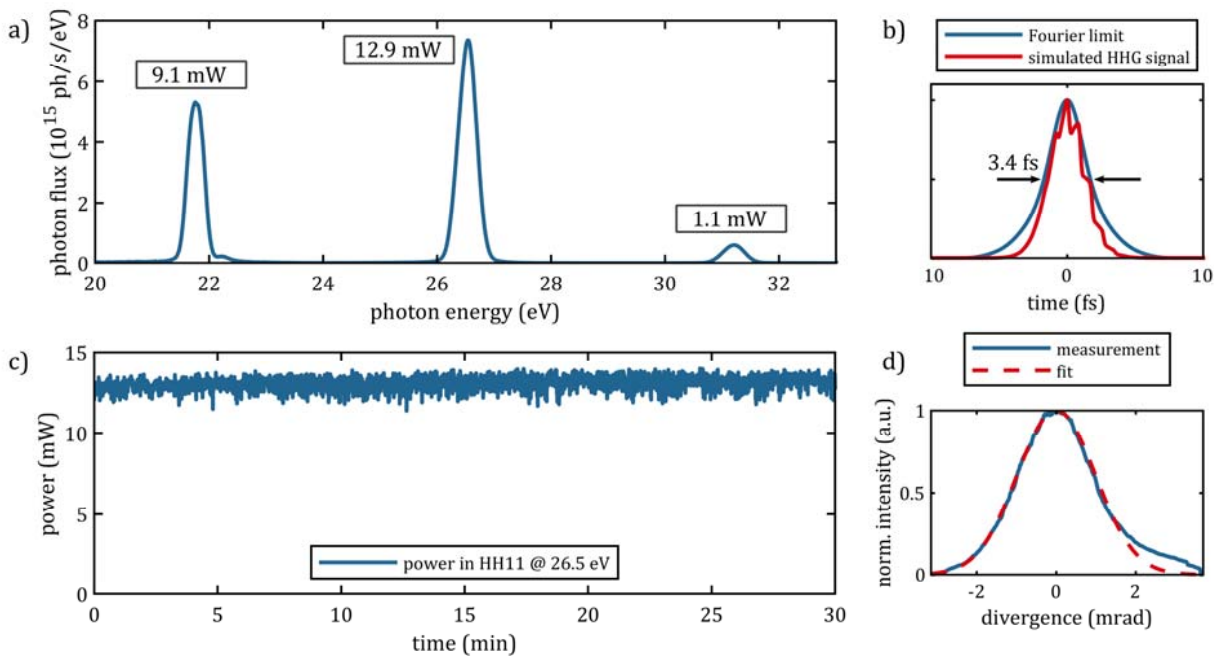

**Fig. 3 High harmonic generation at 1 MHz. a)** Generated high harmonic spectrum using krypton with a backing pressure of 0.4 bar and a 430 μm gas nozzle, together with the corresponding average power in each harmonic line. The highest average power in the 11$^{th}$ harmonic of 12.9 mW corresponds to a photon flux of $3 \cdot 10^{15}$ ph/s. **b)** Fourier limited pulse duration of the 11$^{th}$ harmonic at 26.5 eV of 3.4 fs (FWHM) as well as the simulated high harmonic signal. **c)** Long term measurement of the average power in the 11$^{th}$ harmonic at 26.5 eV over a time period of 30 min every 0.2 sec, showing an rms deviation of 2.6%. **d)** Spatial lineout of the whole XUV beam with a corresponding Gaussian fit.

## Discussion

In conclusion, a high power XUV source via HHG, delivering sub-6 fs pulses at an average power of 12.9 mW in single harmonic line at 26.5 eV and >1 mW at photon energies above 30 eV is demonstrated. This record XUV power together with the cutoff enhancement is enabled by a new-class of driving laser providing a unique combination of a short wavelength (515 nm) and a short pulse duration (18.6 fs) at a record high average power of 51 W. Compared to state-of-the art XUV sources, the provided average power is one order of magnitude higher and the pulse duration is significantly shorter, which represents a major milestone for upcoming applications of coherent XUV radiation in science and technology. This will greatly advance and facilitate XUV application in various fields e.g investigating fastest dynamics using photoelectron emission spectroscopy, coincidence measurements, XUV- ionization and absorption spectroscopy, fluorescence spectroscopy, ultrafast XUV imaging and XUV-pump XUV-probe experiments, among others[10–17,19,20]. Due to the excellent scaling possibilities of energetic ultrafast Yb-based fibre lasers and the components for pulse compression to the kW-level and beyond[24,34,35], even upscaling of the presented approach to the 100 mW level seems possible.

**References**

1. McPherson, A. *et al.* Studies of multiphoton production of vacuum-ultraviolet radiation in the rare gases. *J. Opt. Soc. Am. B* **4**, 595 (1987).
2. Ferray, M. *et al.* Multiple-harmonic conversion of 1064 nm radiation in rare gases. *J. Phys. B At. Mol. Opt. Phys.* **21**, L31–L35 (1988).
3. Popmintchev, T., Chen, M.-C., Arpin, P., Murnane, M. M. & Kapteyn, H. C. The attosecond nonlinear optics of bright coherent X-ray generation. *Nat. Photonics* **4**, 822–832 (2010).
4. Sakdinawat, A. & Attwood, D. Nanoscale X-ray imaging. *Nat. Photonics* **4**, 840–848 (2010).
5. Chang, Z., Corkum, P. B. & Leone, S. R. Attosecond optics and technology: progress to date and future prospects [Invited]. *J. Opt. Soc. Am. B* **33**, 1081 (2016).
6. Krausz, F. & Ivanov, M. Attosecond physics. *Rev. Mod. Phys.* **81**, 163–234 (2009).
7. Constant, E. *et al.* Optimizing High Harmonic Generation in Absorbing Gases: Model and Experiment. *Phys. Rev. Lett.* **82**, 1668–1671 (1999).
8. Takahashi, E., Nabekawa, Y. & Midorikawa, K. Generation of 10-μJ coherent extreme-ultraviolet light by use of high-order harmonics. *Opt. Lett.* **27**, 1920 (2002).
9. Saraceno, C. J., Sutter, D., Metzger, T. & Abdou Ahmed, M. The amazing progress of high-power ultrafast thin-disk lasers. *J. Eur. Opt. Soc. Publ.* **15**, 15 (2019).
10. Keunecke, M. *et al.* Time-resolved momentum microscopy with a 1 MHz high-harmonic extreme ultraviolet beamline. *Rev. Sci. Instrum.* **91**, 063905 (2020).
11. Comby, A. *et al.* Bright, polarization-tunable high repetition rate extreme ultraviolet beamline for coincidence electron–ion imaging. *J. Phys. B At. Mol. Opt. Phys.* **53**, 234003 (2020).
12. Geneaux, R., Marroux, H. J. B., Guggenmos, A., Neumark, D. M. & Leone, S. R. Transient absorption spectroscopy using high harmonic generation: a review of ultrafast X-ray dynamics in molecules and solids. *Philos. Trans. R. Soc. A Math. Phys. Eng. Sci.* **377**, 20170463 (2019).
13. Hütten, K. *et al.* Ultrafast quantum control of ionization dynamics in krypton. *Nat. Commun.* **9**, 719 (2018).
14. Kfir, O. *et al.* Nanoscale magnetic imaging using circularly polarized high-harmonic radiation. *Sci. Adv.* **3**, eaao4641 (2017).
15. LaForge, A. C. *et al.* Time-resolved quantum beats in the fluorescence of helium resonantly excited by XUV radiation. *J. Phys. B At. Mol. Opt. Phys.* **53**, 244012 (2020).
16. Rothhardt, J. *et al.* Lifetime measurements of ultrashort-lived excited states in Be-like ions. *X-Ray Spectrom.* **49**, 165–168 (2020).
17. González-Castrillo, A., Martín, F. & Palacios, A. Quantum state holography to reconstruct the molecular wave packet using an attosecond XUV–XUV pump-probe technique. *Sci. Rep.* **10**, 12981 (2020).
18. Corkum, P. B. & Krausz, F. Attosecond science. *Nat. Phys.* **3**, 381–387 (2007).
19. Lépine, F., Ivanov, M. Y. & Vrakking, M. J. J. Attosecond molecular dynamics: fact or fiction? *Nat. Photonics* **8**, 195–204 (2014).
20. Siegrist, F. *et al.* Light-wave dynamic control of magnetism. *Nature* **571**, 240–244 (2019).
21. Lee, J., Carlson, D. R. & Jones, R. J. Optimizing intracavity high harmonic generation for XUV fs frequency combs. *Opt. Express* **19**, 23315 (2011).
22. Cingöz, A. *et al.* Direct frequency comb spectroscopy in the extreme ultraviolet. *Nature* **482**, 68–71 (2012).

23. Hädrich, S. *et al.* High photon flux table-top coherent extreme-ultraviolet source. *Nat. Photonics* **8**, 779–783 (2014).
24. Müller, M. *et al.* 1 kW 1 mJ eight-channel ultrafast fiber laser. *Opt. Lett.* **41**, 3439 (2016).
25. Porat, G. *et al.* Phase-matched extreme-ultraviolet frequency-comb generation. *Nat. Photonics* **12**, 387–391 (2018).
26. Klas, R. *et al.* Table-top milliwatt-class extreme ultraviolet high harmonic light source. *Optica* **3**, 1167 (2016).
27. Comby, A. *et al.* Cascaded harmonic generation from a fiber laser: a milliwatt XUV source. *Opt. Express* **27**, 20383 (2019).
28. Shiner, A. D. *et al.* Wavelength Scaling of High Harmonic Generation Efficiency. *Phys. Rev. Lett.* **103**, 073902 (2009).
29. Kazamias, S. *et al.* Pressure-induced phase matching in high-order harmonic generation. *Phys. Rev. A* **83**, 063405 (2011).
30. Chang, Z. *Fundamentals of Attosecond Optics*. (CRC Press, 2016). doi:10.1201/b10402
31. Wang, H. *et al.* Bright high-repetition-rate source of narrowband extreme-ultraviolet harmonics beyond 22 eV. *Nat. Commun.* **6**, 7459 (2015).
32. Pronin, O. *et al.* Ultrabroadband efficient intracavity XUV output coupler. *Opt. Express* **19**, 10232 (2011).
33. Klas, R., Kirsche, A., Tschernajew, M., Rothhardt, J. & Limpert, J. Annular beam driven high harmonic generation for high flux coherent XUV and soft X-ray radiation. *Opt. Express* **26**, 19318 (2018).
34. Hädrich, S. *et al.* Scalability of components for kW-level average power few-cycle lasers. *Appl. Opt.* **55**, 1636 (2016).
35. Rothhardt, J. *et al.* 100 W average power femtosecond laser at 343 nm. *Opt. Lett.* **41**, 1885 (2016).